\documentclass[prc,aps,twocolumn,showpacs]{revtex4}
\usepackage{amsmath,amssymb}
\usepackage{graphicx}
\def\pcomma{$^,$}
\def\pkent{$^1$}
\def\pglsg{$^2$}
\def\pmainz{$^3$}
\def\pbonn{$^4$}
\def\pmit{$^{5}$}
\def\pdubna{$^6$}
\def\ppavia{$^{7}$}
\def\pgwu{$^{8}$}
\def\plpi{$^{9}$}
\def\phallifax{$^{10}$}
\def\pbasel{$^{11}$}
\def\ptomsk{$^{12}$}
\def\pedinb{$^{13}$}
\def\pinr{$^{14}$}
\def\psackv{$^{15}$}
\def\pregina{$^{16}$}
\def\pzagreb{$^{17}$}
\def\pamherst{$^{18}$}
\def\pbochum{$^{19}$}
\def\pucla{$^{20}$}
\begin{document}

\title{\boldmath
Measurement of the transverse target and beam-target asymmetries in $\eta$ meson photoproduction at MAMI }

\author{
C.S.~Akondi\pkent, 
J.~R.~M.~Annand\pglsg,
H.~J.~Arends\pmainz,
R.~Beck\pbonn,
A.~Bernstein\pmit,
N.~Borisov\pdubna,
A.~Braghieri\ppavia,
W.~J.~Briscoe\pgwu,
S.~Cherepnya\plpi,
C.~Collicott\phallifax,
S.~Costanza\ppavia,
E.~J.~Downie\pmainz\pcomma\pgwu,
M.~Dieterle\pbasel,
A.~Fix\ptomsk,
L.~V.~Fil'kov\plpi,
S.~Garni\pbasel,
D.~I.~Glazier\pedinb\pcomma\pglsg,
W.~Gradl\pmainz,
G.~Gurevich\pinr,
P.~Hall Barrientos\pedinb,
D.~Hamilton\pglsg,
D.~Hornidge\psackv,
D.~Howdle\pglsg,
G.~M.~Huber\pregina,
V.~L.~Kashevarov\pmainz\pcomma\plpi\footnote[1]{Electronic address: kashev@kph.uni-mainz.de},
I.~Keshelashvili\pbasel,
R.~Kondratiev\pinr,
M.~Korolija\pzagreb,
B.~Krusche\pbasel,
A.~Lazarev\pdubna,
V.~Lisin\pinr,
K.~Livingston\pglsg,
I.~J.~D.~MacGregor\pglsg,
J.~Mancel\pglsg,
D.~M.~Manley\pkent,
P.~Martel\pamherst\pcomma\pmit,
E.~F.~McNicoll\pglsg,
W.~Meyer\pbochum,
D.~Middleton\psackv\pcomma\pmainz,
R.~Miskimen\pamherst,
A.~Mushkarenkov\ppavia\pcomma\pamherst,
B.~M.~K.~Nefkens\pucla\footnote[2]{deceased},
A.~Neganov\pdubna,
A.~Nikolaev\pbonn,
M.~Oberle\pbasel,
M.~Ostrick\pmainz\footnote[3]{Electronic address: ostrick@kph.uni-mainz.de},
H.~Ortega\pmainz,
P.~Ott\pmainz,
P.~B.~Otte\pmainz,
B.~Oussena\pmainz\pcomma\pgwu,
P.~Pedroni\ppavia,
A.~Polonski\pinr,
V.~V.~Polyanski\plpi,
S.~Prakhov\pucla,
G.~Reicherz\pbochum,
T.~Rostomyan\pbasel,
A.~Sarty\phallifax,
S.~Schumann\pmainz\pcomma\pmit,
O.~Steffen\pmainz,
I.~I.~Strakovsky\pgwu,
Th.~Strub\pbasel,
I.~Supek\pzagreb,
L.~Tiator\pmainz,
A.~Thomas\pmainz,
M.~Unverzagt\pmainz,
Yu.~A.~Usov\pdubna,
D.~P.~Watts\pedinb,
D.~Werthm\"uller\pbasel,
L.~Witthauer\pbasel,
and M.~Wolfes\pmainz
\\
(A2 Collaboration at MAMI)
\vspace*{0.1in}
}

\affiliation{
\pkent Kent State University, Kent, Ohio 44242-0001, USA}

\affiliation{
\pglsg SUPA School of Physics and Astronomy, University of Glasgow, Glasgow G12 8QQ,
United Kingdom}

\affiliation{
\pmainz Institut f\"ur Kernphysik, Johannes Gutenberg-Universit\"at Mainz,
D-55099 Mainz, Germany}

\affiliation{
\pbonn Helmholtz-Institut f\"ur Strahlen- und Kernphysik, University of Bonn,
D-53115 Bonn, Germany}

\affiliation{
\pmit Massachusetts Institute of Technology, Cambridge, Massachusetts 02139, USA}

\affiliation{
\pdubna Joint Institute for Nuclear Research, 141980 Dubna, Russia}

\affiliation{
\ppavia INFN Sezione di Pavia, I-27100 Pavia, Italy}

\affiliation{
\pgwu The George Washington University, Washington, DC 20052-0001, USA}

\affiliation{
\plpi Lebedev Physical Institute, 119991 Moscow, Russia}

\affiliation{
\phallifax Department of Astronomy and Physics, Saint Mary’s University,
Halifax, Nova Scotia B3H 3C3, Canada}

\affiliation{
\pbasel Departement f\"ur Physik, University of Basel, CH-4056 Basel, Switzerland}

\affiliation{
\ptomsk Laboratory of Mathematical Physics, Tomsk Polytechnic University, Tomsk, Russia}

\affiliation{
\pedinb SUPA School of Physics, University of Edinburgh, Edinburgh EH9 3JZ, United Kingdom}

\affiliation{
\pinr Institute for Nuclear Research, 125047 Moscow, Russia}

\affiliation{
\psackv Mount Allison University, Sackville, New Brunswick E4L 1E6, Canada}

\affiliation{
\pregina University of Regina, Regina, Saskatchewan S4S 0A2, Canada}

\affiliation{
\pzagreb Rudjer Boskovic Institute, HR-10000 Zagreb, Croatia}

\affiliation{
\pamherst University of Massachusetts, Amherst, Massachusetts 01003, USA}

\affiliation{
\pbochum Institut f\"ur Experimentalphysik, Ruhr-Universit\"at , D-44780 Bochum, Germany}

\affiliation{
\pucla University of California Los Angeles, Los Angeles, California 90095-1547, USA}

\date{today}

\begin{abstract}
We present new data for the transverse target asymmetry $T$ and the very first data for the 
beam-target asymmetry $F$ in the $\vec \gamma \vec p\to\eta p$ reaction up to a center-of-mass energy of $W=1.9$~GeV. 
The data were obtained with the Crystal-Ball/TAPS  detector setup at the Glasgow tagged photon facility of 
the Mainz Microtron MAMI. All existing model predictions fail to reproduce the new data indicating
a significant impact on our understanding of the underlying dynamics of $\eta$ meson photoproduction.
The peculiar nodal structure observed in existing $T$ data close to threshold is not confirmed. 
\end{abstract}


\maketitle

The electromagnetic production of $\eta$ mesons is a selective probe to study resonance 
excitations of the nucleon ($N^{\star}$) for several reasons. Firstly, due to the isoscalar 
nature of the $\eta$ meson, $\Delta^{\star}$ excitations with isospin $I=3/2$ do not contribute 
to the $\gamma N \to \eta N$ reactions. Secondly, due to the smallness of the $\eta NN$ coupling,
non-resonant parts of the scattering amplitudes are strongly suppressed. Therefore, in contrast to 
photoproduction of pions, the dynamics is dominated by resonance excitations.
The photoproduction of $\eta$ mesons is part of dedicated baryon resonance programs at 
MAMI, ELSA and JLab and precision data on unpolarized cross sections and single-spin
observables have already been obtained (see e.g. \cite{Crede2013} for a review). Preliminary 
results for double-spin observables from ELSA and JLab similar to the data presented in this 
letter, have been presented recently (see e.g. \cite{CBELSA, CLAS}).
Analyses of these data have been performed within single- and multi-channel 
isobar models \cite{MAID, MAIDr, BnGa10, BnGa11} and coupled-channel 
approaches \cite{Giessen2012, Kamano2013}. Furthermore, a partial-wave analysis has been performed 
within the SAID formalism \cite{McNicoll2010}.
All these analyses agree in the fact that the low-energy behavior of the
$\eta$ production process is governed by the $E_{0+}$ multipole amplitude, which is populated by 
the $N^{\star}(1535)1/2^-$ resonance. Higher mass $1/2^-$ resonances also appear to couple 
strongly to the $\eta N$ channel. Other resonances, with
a small branching fraction to $ \eta N$, can be identified by exploiting interference with the dominant 
$E_{0+}$ amplitude in single- and double-spin observables. 
The beam asymmetry $\Sigma $, measured with a linearly polarized photon beam \cite{Ajaka1998, Elsner2007}, 
and the transversely polarized target asymmetry $T$ \cite{Bock1998} are particularly sensitive to
an interference of $s$- and $d$-wave amplitudes. A model independent analysis in the threshold region 
allowed for the determination of parameters of the $N^{\star}(1520)3/2^-$ resonance \cite{Tiator1999f}
and its contribution to $\eta$ photoproduction. 
However, the target asymmetry data of \cite{Bock1998} did not fit into this overall picture. 
The observed nodal structure in the threshold region could not be described by {\it any} 
reaction model using Breit-Wigner shapes for the parametrization of nucleon resonance 
contributions. 
The model independent, truncated multipole analysis \cite{Tiator1999f}
showed that this feature enforced a large and
rapidly varying phase between the $E_{0+}$ and the $E_{2-}$, $M_{2-}$ multipoles.
This phase was later supported by a measurement of the proton recoil
polarization in the $p(e,e' \vec{p})\eta$ reaction \cite{Merkel2007}.
However, such a strong phase motion is not possible between amplitudes dominated by two 
Breit-Wigner resonances with very close pole positions, the $N^{\star}(1535)1/2^-$ and the $N^{\star}(1520)3/2^-$.
Since the original $T$ data \cite{Bock1998} had quite
significant uncertainties, a more precise measurement of this observable in order to
confirm or refute the nodal structure was highly desirable.

A second exciting observation was a narrow structure in the excitation
function of $\eta$ photoproduction off the neutron at 
$W=1670$~MeV \cite{Kuznetsov2007a, Miyahara2007, Jaegle2011a, Werthmuller2013}. 
The position coincides with a dip observed in the $\gamma p \to \eta p$ 
total cross section \cite{McNicoll2010}.
The interpretations discussed in the literature include new narrow resonances, an interference between 
$1/2^-$ resonances, or coupled channel effects due to the opening of $K\Lambda$ and $K\Sigma$ channels.

In this letter, we report a new, high-statistics measurement of $\eta$ photoproduction from transversely polarized 
protons. The differential cross section is given by
\begin{equation}\label{crosssection}
  \frac{d\sigma}{d\Omega} = \frac{d\sigma_0}{d\Omega} \left( 1 + P_T \sin\phi  \; T  + h P_{\odot} P_T \cos \phi  \; F \right). 
\end{equation}
Here $P_{\odot}$ and $P_T$  denote the degree of circular beam and transverse target polarization,  $h= \pm 1$ is the beam 
helicity and $\phi$ is the azimuthal angle of the target polarization vector in 
a coordinate frame fixed to the reaction plane with $ \hat z = \vec p_{\gamma}/|\vec p_{\gamma}|$, 
$\hat y = \vec p_{\gamma} \times \vec p_{\eta}/|\vec p_{\gamma} \times \vec p_{\eta}|$ and 
$\hat x = \hat y \times \hat z$.

The experiment was performed at the MAMI C accelerator in Mainz\,\cite{MAMIC} using the
Glasgow-Mainz tagged photon facility\,\cite{TAGGER}. In the present measurement, a
longitudinally polarized electron beam with an energy of 1557 MeV and a polarization degree of
80\% was used. The tagged photon beam covers the energy 
range from 700 to 1450 MeV. The longitudinal polarization of electrons is transferred to circular
polarization of the photons during the bremsstrahlung process in a radiator.
The degree of circular polarization depends on the photon energy and ranged from 65\%
at 700 MeV to 78\% at 1450 MeV \cite{Olsen}.
The reaction $\gamma p\to \eta p$ was measured using the Crystal Ball
(CB)\,\cite{CB} as the central spectrometer and TAPS \,\cite{TAPS} as a forward
spectrometer.
The combined CB/TAPS detection system covers $97\%$ of the full solid angle. 
More details on the energy and angular resolution of CB and TAPS are given
in Ref.\,\cite{Setup}.

The experiment requires transversely polarized protons, which were provided by a
frozen-spin butanol ($\mathrm{C_4H_9OH}$) target.
A specially designed $\mathrm{^3He/^4He}$ dilution refrigerator was built in order to maintain 
a temperature of 25 mK during the measurements. 
For transverse polarization, a 4-layer saddle coil was installed as the holding magnet,
which operated at a current of 35 A, corresponding to a field of 0.45 Tesla.
The target container, length 2 cm and diameter 2 cm, was filled with 2-mm diameter butanol
spheres with a packing fraction (filling factor) of around $60\%$.
The average proton polarization during the beam time periods May-June 2010 and April 2011
was $70\%$ with relaxation times of around 1500 hours. The target polarization was measured
at the beginning and the end of each data taking period. In order
to reduce the systematic errors, the direction of the target polarization vector was regularly 
reversed during the experiment.
More details about the construction and operation of the target are given in Ref.\,\cite{Thomas}.

\begin{figure}
\begin{center}
\resizebox{0.48\textwidth}{!}{%
\includegraphics{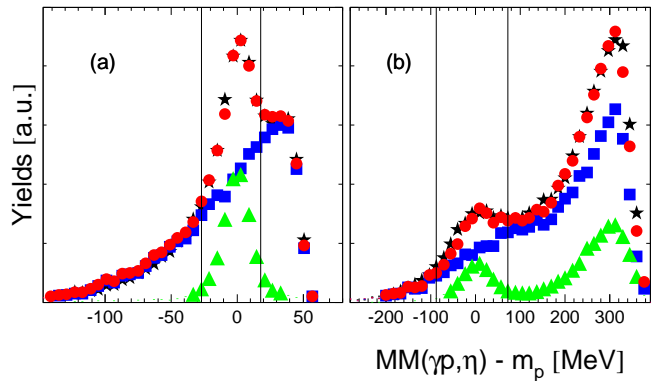}}
\caption{(Color online) Two typical examples of the carbon background subtraction,
corresponding to an $\eta$ meson polar angle around $90^o$ and photon beam energies of 
785 MeV (a) and 1350 MeV (b).    
The $MM(\gamma p, \eta)$ missing mass distributions obtained with butanol are shown as black stars. 
The green triangles and blue squares are the 
distributions obtained with hydrogen and carbon target scaled to fit the butanol data. 
The red circles are the sum of the blue and green distributions.}
\label{fig1}
\end{center}
\end{figure}
\begin{figure}
\begin{center}
\resizebox{0.48\textwidth}{!}{%
\includegraphics{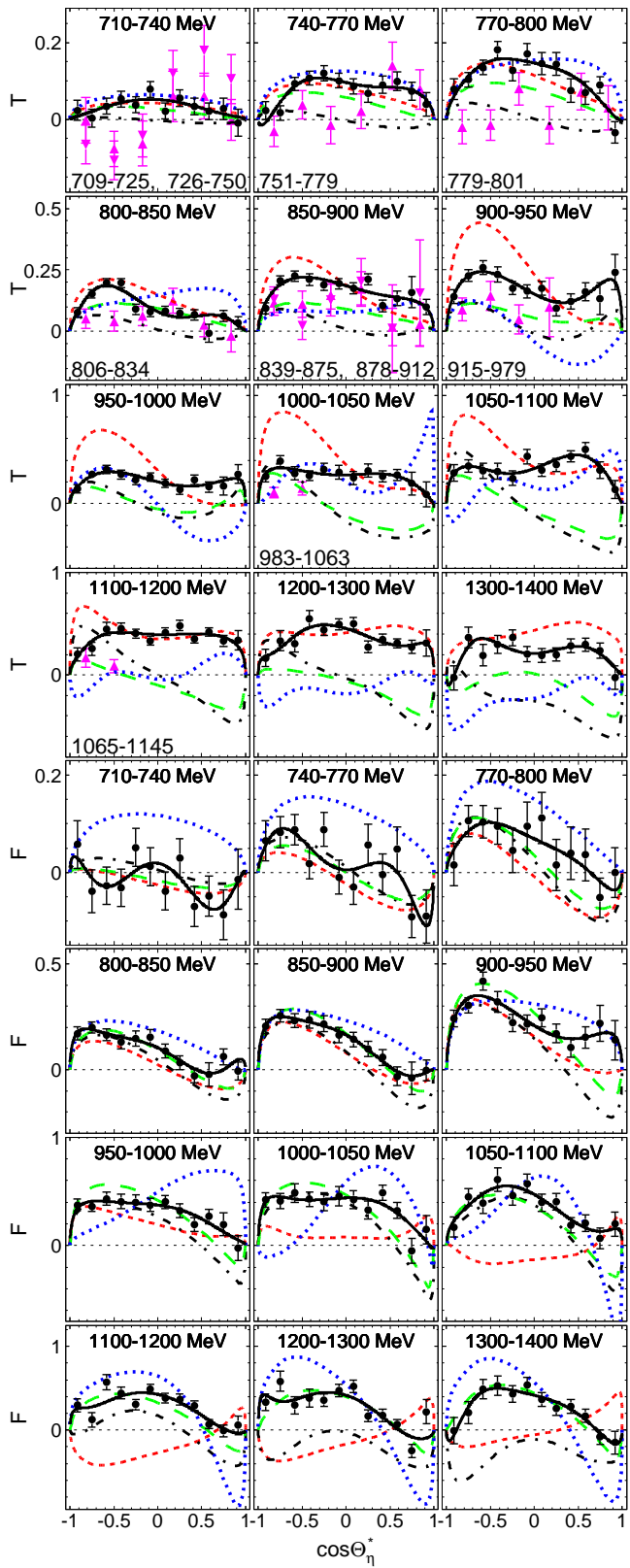}}
\caption{(Color online) $T$ and $F$ asymmetries.
The new results with statistical uncertainties (black circles) are compared to 
existing data from Bonn \cite{Bock1998} (magenta triangles) and existing PWA 
predictions (red dashed: $\eta$-MAID \,\cite{MAID}, 
green long-dashed: Giessen model\,\cite{Giessen2012}, 
black dashed-dotted: BG2011-02 \,\cite{BnGa11},
blue dotted: SAID GE09 \,\cite{McNicoll2010}). 
The result of our Legendre fit is shown by the black curves, Eq.~\ref{NF}.
The energy labels on the top of each panel indicate the photon energy bins
for our data. The values at the bottom give the corresponding bins of \cite{Bock1998}.}
\label{fig2}
\end{center}
\end{figure}
The mesons were identified via the $\eta \to 2 \gamma$ or $\eta \to 3 \pi^0 \to 6 \gamma$ decays. 
Selections on the $2 \gamma$, or $6 \gamma$, invariant 
mass distributions and on the missing mass, $MM(\gamma p, \eta)$, calculated from the initial state and
the reconstructed $\eta$ meson, allowed for a clean identification of the reaction.
In principle, the observables $T$ and $F$ in Eq.~\ref{crosssection} can be determined in each energy 
and angular bin as count rate asymmetries from the number $N^{\pm}$ of reconstructed 
$\vec \gamma \vec p \to \eta p$ events with different 
orientations of target spin and beam helicity:
\begin{equation}\label{NT}
T =\frac{1}{P_T |\sin\phi|}\,\frac{N^{\pi = +1}-N^{\pi = -1}}{N^{\pi = +1}+N^{\pi = -1}}\,,        
\end{equation}
\begin{equation}\label{NF}
F =\frac{1}{P_T |\cos\phi|}\,\frac{1}{P_\odot}\,
\frac{N^{\sigma = +1}-N^{\sigma = -1}}{N^{\sigma = +1}+N^{\sigma = -1}}\,,      
\end{equation}  
where $\pi = \vec p_T \cdot \hat y/|\vec p_T \cdot \hat y| = \pm 1$ denotes the orientation of the target 
polarization vector 
$\vec p_T$ relative to the normal of the production plane and, in the case of the $F$ asymmetry,  
$\sigma = h \; \vec p_T \cdot \hat x/|\vec p_T \cdot \hat x| = \pm 1$ is given by the product of the beam helicity $h$ 
and the orientation of $\vec p_T$ relative to the $\hat x$ axis. In these asymmetries, systematic uncertainties 
related to the reconstruction efficiency, the total photon flux normalization and the target filling factor cancel.
However, using butanol as target material has an essential consequence because of the 
\begin{figure*}
\begin{center}
\resizebox{1.0\textwidth}{!}{%
\includegraphics{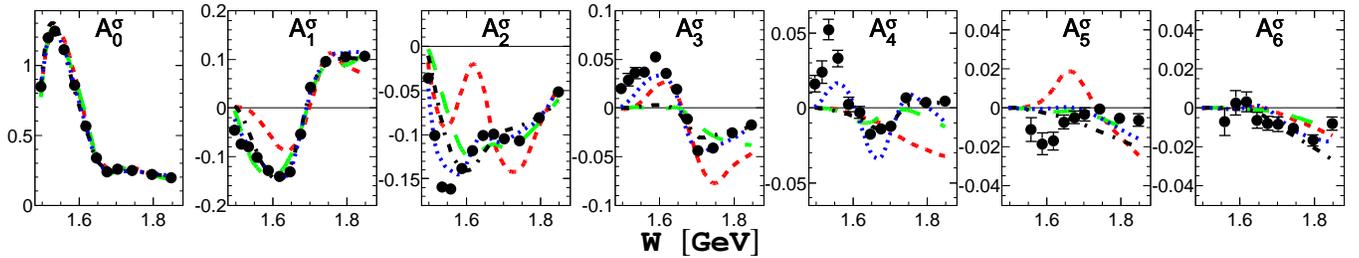}}
\caption{(Color online) Legendre coefficients in $[\mu b/sr]$ up to $\ell_{\mathrm{max}} = 3$ from 
our fits to the differential
cross section \cite{McNicoll2010} as function of the center-of-mass energy $W$.
Notations for the curves are the same as in Fig.\,\ref{fig2}}.
\label{fig3}
\end{center}
\end{figure*}
\begin{figure*}
\begin{center}
\resizebox{1.0\textwidth}{!}{%
\includegraphics{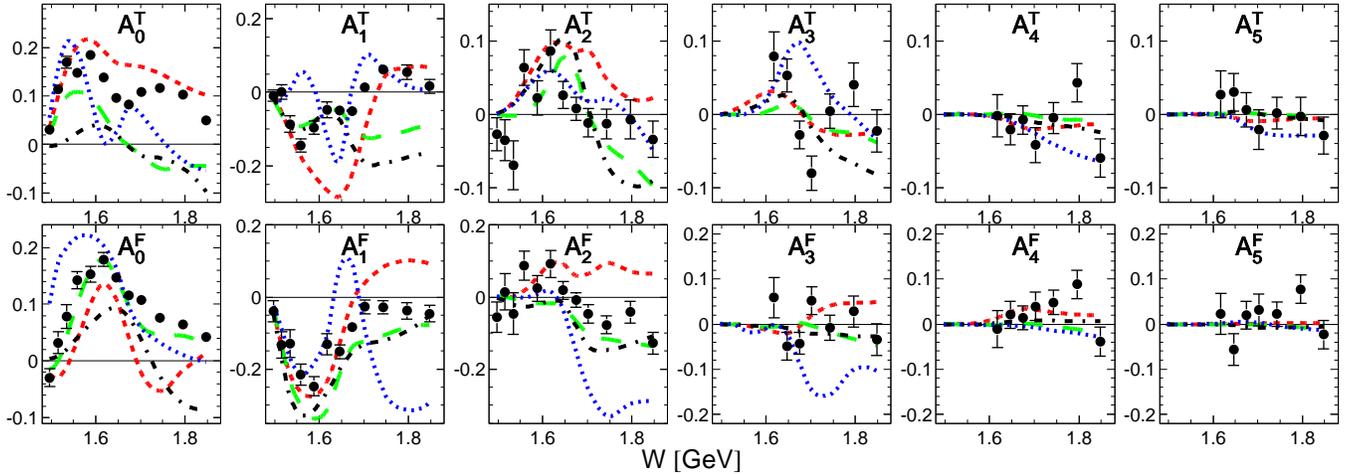}}
\caption{(Color online) Legendre coefficients $[\mu b/sr]$ up to $\ell_{\mathrm{max}}= 3$ from our 
fits to the product of the 
new asymmetries with the differential cross section from \cite{McNicoll2010}:
$T d\sigma/d\Omega$ (upper row) and $F d\sigma/d\Omega$ (lower row).
Notations for the curves are the same as in Fig.\,\ref{fig2}}.
\label{fig4}
\end{center}
\end{figure*}
background coming from quasi-free reactions on $\mathrm{^{12}C}$ and $\mathrm{^{16}O}$ nuclei. In the numerator 
of Eqs.~\ref{NT} and \ref{NF}, this background cancels because the nucleons bound in 
$\mathrm{^{12}C}$ or $\mathrm{^{16}O}$ are
unpolarized. However, in order to determine the denominator, this contribution has 
to be taken into account.  The detection of recoil
protons and the requirement of co-planarity of the incoming photon and the outgoing hadrons 
already suppress this background significantly. 
The residual background has to be subtracted. In order to do this, the shape of the 
missing mass distribution $MM(\gamma p, \eta)$
was determined for $\eta$ photoproduction on a pure carbon and a 
liquid hydrogen target. These templates were then fitted to the butanol data.
Since the magnitude and the shape of the background depend on the initial beam
energy and momenta of the final particles, the background subtraction procedure was
performed for each energy and angular bin. 
This procedure is illustrated in Fig.\,\ref{fig1} for
two different examples at low and at high photon energy. Missing mass
spectra for the
reaction $\gamma p\to \eta p$ with the butanol target are shown in Fig.\,\ref{fig1}
(a) and (b) by the black stars. Spectra measured with the hydrogen and carbon
targets are presented on the same plots by the green triangles and the blue squares
correspondingly. Their absolute values were fitted to the butanol data with a typical 
reduced $\chi^2$ between $0.7$ and $1.5$. The red
circles, representing the sum of the hydrogen and carbon contribution, are the result of
this fit. The signal is located around $MM(\gamma p, \eta) = m_p$. At higher missing masses 
and higher photon energies additional background from multi-meson final states is observed. 
The number of signal events 
was determined in the regions between the vertical solid lines, which were selected 
to optimize the signal-to-background ratio and to remove 
background from multi-meson production off polarized protons. 

The systematic uncertainty is dominated by the determination of the degree of proton polarization (4\%), 
the degree of photon beam polarization (2\%), and the background subtraction procedure (3-4\%).
By adding all contributions in quadrature, a total systematic uncertainty of less than 6\% is obtained.

Fig.\,\ref{fig2} shows our results for $T$ and $F$ asymmetries together with previous data
for $T$ \,\cite{Bock1998} and various theoretical predictions  
\,\cite{MAID,BnGa10, McNicoll2010,Giessen2012}
for different bins in the incoming photon energy as function 
of the $\eta$ meson polar angle in the center-of-mass system, $\theta_{\eta}^*$.
The main inconsistencies with the existing data \cite{Bock1998} are in the near 
threshold region. Here, our results do not confirm the observed nodal structure in the angular 
dependence of the $T$ asymmetry and solve the long-standing question related to the relative phase 
between $s$- and $d$-wave amplitudes. Our data do not require any additional phaseshift beyond a 
Breit-Wigner parametrization of resonances. 
This important conclusion is corroborated by preliminary data from ELSA \cite{CBELSA}. 
At higher energies, all existing theoretical 
predictions of both $T$ and $F$ 
are in poor agreement among themselves and with our experimental data, even though they describe the 
unpolarized differential cross sections well. The new data will therefore have a significant 
impact on the partial-wave structure of all models. 

Furthermore, we present a fit of our cross section data \cite{McNicoll2010} and the 
new polarization measurements based on 
an expansion in terms of Legendre polynomials truncated to a maximum orbital angular momentum $\ell_{\mathrm{max}}$:
\begin{eqnarray}
\frac{d\sigma}{d\Omega} &=&
\sum\limits_{n=0}^{2 \ell_{\mathrm{max}}} A^{\sigma}_{n}P_{n}(\cos\Theta_{\eta})  \\\label{LegPol2}
T (F) \;\; \frac{d\sigma}{d\Omega} &=& \sin\Theta_{\eta}
\sum\limits_{n=0}^{2 \ell_{\mathrm{max}} - 1} A^{T(F)}_{n}P_{n}(\cos\Theta_{\eta}). \label{LegPol1}
\end{eqnarray} 
The spin-dependent cross sections, $T d\sigma/d\Omega$ and $F d\sigma/d\Omega$,
were obtained by multiplying the measured asymmetries with our results for the differential cross sections 
\cite{McNicoll2010}.

The results for the Legendre coefficients 
are presented in Figs.\,\ref{fig3}-\ref{fig4} together with the corresponding model calculations.
For the differential cross section a truncation to $\ell_{\mathrm{max}}=2$ ($d$ waves, $A^{\sigma}_4$) 
is sufficient below $W = 1.6$~GeV  
and to $\ell_{\mathrm{max}} = 3$ ($f$ waves, $A^{\sigma}_6$) above $W = 1.6$~GeV. 
Additional higher order terms do not improve the quality of the fit.
For the new spin-dependent cross sections 
a truncation to $pd$ interferences ($A^{T/F}_2$) below $W = 1.6$~GeV and $df$ interferences ($A^{T/F}_4$) 
above  $W = 1.6$~GeV is sufficient.
The result of the Legendre fits is demonstrated in Fig.~\ref{fig2} by the black solid line.
The models \cite{BnGa11,McNicoll2010,Giessen2012} that have been fitted to the differential cross section 
from \cite{McNicoll2010} 
are, as expected, also in agreement with the coefficients in Fig.~\ref{fig3}. Some deviations can be 
observed in $A^{\sigma}_5$, which is dominated by an interference between $d$ and $f$ waves. 
Despite of this agreement, the corresponding predictions for the coefficients 
$A^{T}_{n}$ and $A^F_{n}$ do not agree with our results for all values of $n$ even 
at low energies. The impact of the new data is therefore not restricted to a single partial-wave amplitude but 
that all $s$-, $p$-, and $d$-wave amplitudes will be affected
in future partial-wave analyses. This is in particular the case in the energy region around $W=1670$~MeV, where the
narrow structure in $\eta$ production off neutrons is observed. 
A recent analysis in the framework of the
Bonn-Gatchina analysis claimed that the structure can be completely explained by 
an interference of the $N^{\star}(1535)1/2^-$ and $N^{\star}(1650)1/2^-$ resonances
the  without adding additional contributions from narrow states \cite{Anisovich:2014}. 
The Giessen-model \cite{Giessen2012}  also  explains the structure by an interference within 
the $E_{0+}$ partial wave. Here, the nature of the interference is related to coupled channel effects 
due to the opening of $K$-hyperon channels.
However, as shown in Fig.\ref{fig2}, the predictions of both models for the
target asymmetry in this energy region disagree completely with the new data, in shape as
well as in sign. 
Consequently, such interpretations must be still taken with care and it has to be seen whether 
it is possible to refit these models including the new $T$ data.    

In summary, we have presented new experimental results for the target asymmetry $T$ and 
very first data on the transverse beam-target observable $F$ for the $\vec \gamma \vec p\to\eta p$ reaction. 
The data solve a long-standing problem related the angular dependence of older $T$ data close to threshold.
The unexpected relative phase motion between $s$- and $d$-wave amplitudes required by 
the old data is not confirmed. A Legendre decomposition of the new results shows 
the sensitivity to small partial-wave contributions. There is no evidence for any narrow structure. 
However, all existing solutions from various partial wave analyses fail to reproduce the new data. 
We therefore expect a significant impact on future analyses and on our  
understanding of the dynamics of $\eta$ photoproduction.

The authors wish to acknowledge the excellent support of the accelerator group of MAMI.
This work was supported by the Deutsche Forschungsgemeinschaft (SFB 443, SFB 1044), the
European Community Research Activity under the FP7 program (Hadron Physics, Contract No.
227431), Schweizerischer Nationalfonds, the UK Sciences and Technology Facilities
Council (STFC 57071/1, 50727/1), U.S. DOE, U.S. NSF, and NSERC (Canada), the Dynasty foundation and the MSE
Program ``Nauka'' (Contract No. 1.604.2011).


\end{document}